\documentclass{aastex}          
\usepackage{spr-astr-addons}   
\usepackage{natbib}
\begin{document}
\title{Experiments to investigate the effects of radiative cooling on plasma jet collimation}

\shorttitle{Radiative cooling and jet collimation}
\shortauthors{Gregory et al.}

\author{C. D. Gregory\altaffilmark{1}} \affil{Laboratoire pour l'Utilisation des Lasers Intenses, UMR7605, CNRS - CEA - Universit\'e, Paris VI - Ecole Polytechnique, 91128 Palaiseau Cedex, France}  \email{cdg104@york}
 
\author{A. Diziere}  \affil{Laboratoire pour l'Utilisation des Lasers Intenses, UMR7605, CNRS - CEA - Universit\'e, Paris VI - Ecole Polytechnique, 91128 Palaiseau Cedex, France} 

\author{H. Aoki} \affil{Institute of Laser Engineering, Osaka University, Suita, Osaka, 565-0871, Japan} 

\author{H. Tanji} \affil{Institute of Laser Engineering, Osaka University, Suita, Osaka, 565-0871, Japan}

\author {T. Ide} \affil{Institute of Laser Engineering, Osaka University, Suita, Osaka, 565-0871, Japan}

\author{M. Besio} \affil{Dipartmento di Fisica ``G. Occhialini'' University of Milano-Bicocca, Milan, Italy}

\author{S. Bouquet\altaffilmark{2}} \affil{CEA/DIF/DPTA BP 12, 91680 Bruyers-le-Chatel, France}

\author{E. Falize\altaffilmark{2}} \affil{CEA/DIF/DPTA BP 12, 91680 Bruyers-le-Chatel, France}

\author{B. Loupias} \affil{Laboratoire pour l'Utilisation des Lasers Intenses, UMR7605, CNRS - CEA - Universit\'e, Paris VI - Ecole Polytechnique, 91128 Palaiseau Cedex, France}

\author{C. Michaut} \affil{LUTH, Observatoire de Paris, CNRS, Universit\'e Paris Diderot, Place Jules Janssen, 92190, Meudon, France}

\author{T. Morita} \affil{Institute of Laser Engineering, Osaka University, Suita, Osaka, 565-0871, Japan}

\author{S. A. Pikuz Jr.} \affil{Joint Institute for High Temperatures of RAS, Izhorskaya 13/19, Moscow, 125412, Russia}

\author{A. Ravasio} \affil{Laboratoire pour l'Utilisation des Lasers Intenses, UMR7605, CNRS - CEA - Universit\'e, Paris VI - Ecole Polytechnique, 91128 Palaiseau Cedex, France} 

\author{Y. Kuramitsu} \affil{Institute of Laser Engineering, Osaka University, Suita, Osaka, 565-0871, Japan}

\author{Y. Sakawa} \affil{Institute of Laser Engineering, Osaka University, Suita, Osaka, 565-0871, Japan}

\author{H. Takabe} \affil{Institute of Laser Engineering, Osaka University, Suita, Osaka, 565-0871, Japan}

\author{N. C. Woolsey}
\affil{Department of Physics, University of York, Heslington, YO10 5DD, UK}

\and

\author{M. Koenig}
\affil{Laboratoire pour l'Utilisation des Lasers Intenses, UMR7605, CNRS - CEA - Universit\'e, Paris VI - Ecole Polytechnique, 91128 Palaiseau Cedex, France}

\altaffiltext{1}{Current address: Department of Physics, University of York, Heslington, YO10 5DD, UK}
\altaffiltext{2}{LUTH, Observatoire de Paris, CNRS, Universit\'e Paris Diderot, Place Jules Janssen, 92190, Meudon, France}

\begin{abstract}
Preliminary experiments have been performed to investigate the effects of radiative cooling on plasma jets. Thin (3 $\mu$m - 5 $\mu$m) conical shells were irradiated with an intense laser, driving jets with velocities $>$ 100 km s$^{-1}$. Through the use of different targets materials - aluminium, copper and gold - the degree of radiative losses was altered, and their importance for jet collimation investigated. A number of temporally-resoved optical diagnostics was used, providing information about the jet evolution. Gold jets were seen to be narrower than those from copper targets, while aluminium targets produced the least collimated flows.  
\end{abstract}

\keywords{Young stellar objects; Jets and outflows; Laboratory astrophysics}

\section{Introduction}

Jets from young stellar objects (YSOs) are associated with the accretion phase of stellar evolution, which lasts for around the first $10^5$ years of a young star's life, see \citet{rei:her01} for a review. The jets are seen propagating away from the star at speeds of the order of 500 km s$^{-1}$, with lengths of up to 1 pc, and with aspect ratios (jet length / jet width) of 10 or more. The flows are often seen to terminate in regions of optical emission, and contain a series of bright knots - these regions of emission are known as Herbig-Haro (HH) objects. The relative proximity of these systems means that observations of radiation from HH objects have allowed a large amount of high-quality observational data to be collected. These objects have dynamic time-scales of the order of a decade, and it is thus possible to observe the jet as it evolves. These data, along with theoretical and computational modelling, have lead to significant improvements in the understanding of YSO jets. Despite this, questions still exist surrounding the physics of both the jet launching and propagation, and so the possibility of performing well-designed laboratory simulations that contribute to current level of understanding therefore has potentially large benefits. This experiment represents part of an ongoing campaign \citep{gre:ast08,lou:sup07,gre:ast08b,wau:aje09} to aid in the understanding of aspects of jet propagation physics, and in particular the high degree of collimation over very large distances. There are three main physical processes which are thought to be important: 1/ inertial confinement of the flow by the ambient medium, 2/ radiative cooling of the jet, causing a drop in the internal thermal pressure and a collapse on axis, and 3/ magnetic fields which act to restrict the radial expansion of the charged particles. A number of experiments have taken place in recent years to investigate jet propagation. These have demonstrated the propagation of a jet moving into an ambient medium, either stationary \citep{nic:stu08,fos:hig05, amp:for05} or in the form of a plasma crosswind \citep{leb:jet04}, and an increase in jet collimation due to radiative cooling \citep{shi:exp00,leb:lab02, pur:col10}. The work presented here is aimed at studying the effects of radiative losses on the collimation of plasma jets. Conical targets are irradiated with an intense laser, resulting in jets of high velocity. By choosing different targets materials - here gold, copper and aluminium - the importance of radiative losses can be altered, since higher atomic number materials radiate more efficiently. This radiation cools the jets on axis, and lowers the thermal pressure driving the radial expansion. As a result, more radiative jets are expected to have larger aspect ratio and to be more collimated. Results are given from two experiemental campaigns, the first at the GEKKO XII, 12 beam, 10 kJ, Nd:glass laser system at the Institute for Laser Engineering, Japan. The second experiment took place at the PICO2000 facility at the Ecole Polytechnique, France, which houses a 100 J, picosecond, Nd:glass laser and a 1 kJ, nanosecond, Nd:glass laser. 

\section{Experimental design}
\begin{figure}[ht]
\begin{centering}
\includegraphics[]{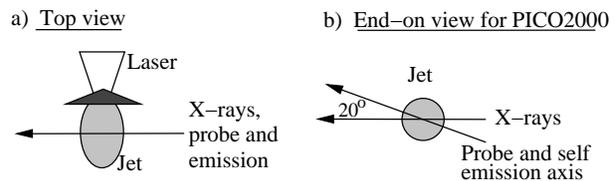}
\caption{The experimental set-up. Panel a) shows the orientation of the optical probe and self-emission diagnostics used in both experiments, as well as the x-ray radiography used only in the PICO200 experiment. The angular separation between the optical and x-ray axes used in the PICO2000 experiment is shown in panel b). This configuration allowed both diagnostics to be used simultaneously. In this view the jet is propagating out of the page.}
\label{fig1}
\end{centering}
\end{figure}

The primary experimental configuration is shown in Figure \ref{fig1}. Five beams of the GEKKO XII laser irradiated the apex of the conical shell, delivering around 500 J of laser light in a 500 ps pulse, at a wavelength of 351 nm. The focal spot size was 600 $\mu$m. The plasma from the rear face of the target is focused on axis due to the conical geometry, and forms a jet. A second, mJ, 527 nm, 20 ns laser pulse probed the system in the direction perpendicular to the plasma flow. The delay between the arrival of the probe beam and the drive lasers was varied up to a maximum of 70 ns. A modified Normarski interferometer analysed the beam, and was detected with a gated optical imager (GOI) with a temporal resolution of 250 ps. In addition, the self emission from the jet at a wavelength of 450nm $\pm 10$ nm was collected and imaged onto two more GOIs, both with a 1.5 ns gate width, and a streak camera with a 28 ns time window. The cone targets were made from either 5 $\mu$m thick gold, 5 $\mu$m thick copper, or 3 $\mu$m thick aluminium. All the targets had a full opening angle of 140$^{\circ}$ and a diameter of around 1.2 mm. 

A second experiment took place at the PICO2000 facility. In this case only copper targets were used, and were irradiated with $\sim 400$ J, 1 ns, 532 nm laser pulses, focused through hybrid phase plates to give a 400 $\mu$m focal spot. A  mJ, 527 nm, 8 ns laser was again used to probe in the direction perpendicular to the jet propagation axis, and was detected with a GOI with a time resolution of 120 ps. Here, no interferometer was used, and the diagnostic was used in a shadowgraphy configuration. A streak camera, with a time window of 50 ns, recorded the self-emission at 450nm $\pm 10$ nm transverse to the jet propgation direction. An additional x-ray radiography diagnostic was implemented for this experiment. Laser pulses of 60 J, 30 ps at 1064 nm, and a focal spot of 50 $\mu$m, irradiated a Ti foil placed 20 mm away from the concial shell. The resulting Ti-k$\alpha$ emission at 4.75 keV was used to radiograph the jets at varying times during their evolution, using imaging plate as the detector. The axis of the x-ray and optical diagnostics were separated by 22 degrees, allowing both diagnostics to be used simultaneaously. 

\section{Results}

\begin{figure*}[ht]
\begin{centering}
\includegraphics[]{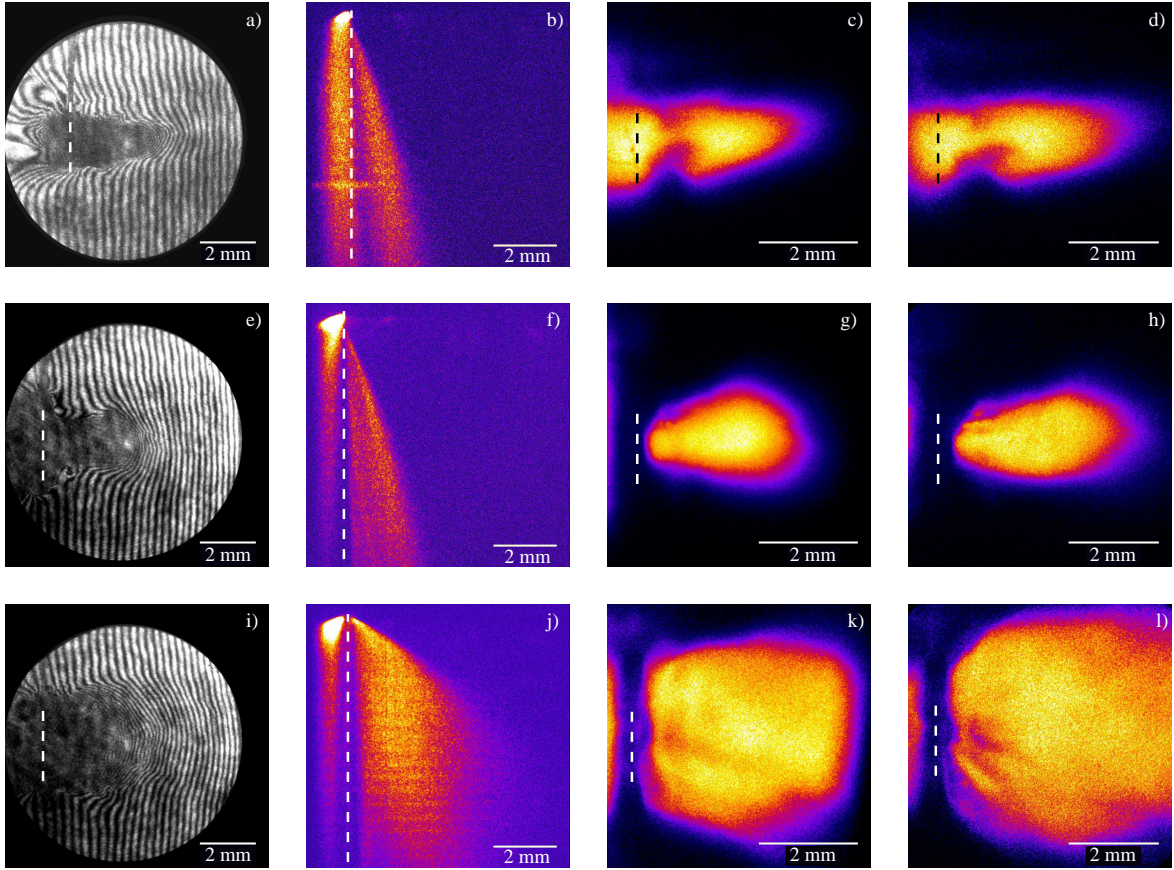}
\caption{The data from the optical diagnostics used in the GEKKO XII experiment. The top, middle and bottom rows represent data for gold, copper, and aluminium targets respectively. The colums, from left to right, show data from the interferometry with a probe delay of 50 ns, the streaked self-emission over the first 28 ns of the jet evolution, the imaged self-emission after 50 ns, and the imaged self-emission after 70 ns. The dashed line indicates the intial target location on the horizontal axis. See text for further details.}
\label{fig2}
\end{centering}
\end{figure*}

Figure \ref{fig2} shows the results of the experiment at the GEKKO XII facility for targets of gold, copper, and aluminium. In all images the laser is incident from the left, and the jet propagates from left to right. The original horizontal postion of the target is indicated by a dashed vertical line. The columns show, respectively, interferograms taken with a probe delay of 50 ns, the streaked self-emission over the first 28 ns of the jet evolution, the imaged self-emission after 50 ns, and the imaged self-emission after 70 ns. In each of the self-emission images, a dark region is seen close to the original target position. This is due to absorption of the emitted light by plasma generated from the target mount. In the case of the gold jet, the interferograms and the self-emission images show a jet with aspect ratio (jet length to jet width) of around 2.5, and the flow appears to be converging onto the axis even at distance far from the original target postion. The initial velocity of the jet can be inferred from the streaked self-emission diagnostic, and is 130 km s$^{-1}$.  The results for the copper targets indicate a jet moving with a velocity of 120 km s$^{-1}$. The reason that this jet has a lower velocity than for gold is likely due to fluctations in the laser energy on target and difficulties with the laser beam alignment process. In this case the jet diverges as it moves away from the target surface, as seen in the self-emission images, which show a flow opening angle of $\sim 30^{\circ}$. In the case of the aluminium cones, the plasma flowing from the target surface moves initially with a much higher velocity of 470 km s$^{-1}$. Due to this the leading edge of the expansion is out of the view window for the interferogram, and for the self-emission image after 70 ns.  

\begin{figure}[ht]
\begin{centering}
\includegraphics[]{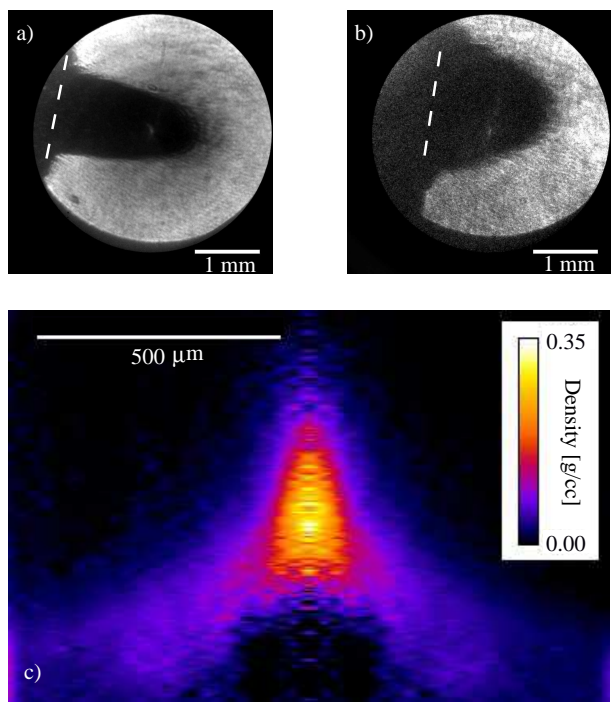}
\caption{Shadowgraphs for a) a concial copper target, and b) a plane copper target, taken with a probe delay of 20 ns during the PICO2000 experiement. Panel c) shows a density profile after 12 ns, inferred from x-ray radiography. In this x-ray image, the jet propagates from bottom to top, and the orginal position of the target is at the bottom edge of the image.}
\label{fig3}
\end{centering}
\end{figure}

Figure \ref{fig3} shows data taken during the PICO2000 experiment. Panel a) shows a shadowgraphic image of a cone target taken after 20 ns. In panel b), a corresponding image is shown for a simple plane foil of 5 $\mu$m thick copper.  The flow from the conical foil is narrower and has propagated a greater distance from the original target position. These data indicate the importance of the target geometry for the formation of jets. Panel c) shows a mass density profile for a jet after 12 ns, inferred form the x-ray radiography.

\section{Discussion}

The data presented in Figures \ref{fig2} and \ref{fig3} show that for both campaigns, the experimental configuration was successful in generating plasma jets. Panels a) and b) of Figures \ref{fig3} indicated the importance of the conical geometry of the target: the cone target produces a significantly more collimated outflow than a simple plane target. The optical and x-ray probes are complimentary - the x-rays are able to diagnose the dense part of the jet close to the target surface, but the more tenous regions of the flow around the jet head are below the sensitivity of the diagnostic. In contrast, the optical probe is unable to penetrate the high-density core of the jet, but is well suited to imaging relatively low plasma densities. The capability of fielding these two diagnostics on a single shot is therefore beneficial, and allows a more complete picture of the dynamics of the jet to be built.

The results shown in Figure \ref{fig2} suggest that the jet collimation is increased for higher-atomic number targets. Gold jets are narrower than those of copper, and the least collimated flows are from the aluminium targets. Heavier materials radiate more efficiently, and the results are consistent with computer simulations \citep{miz:num02}, which have shown that increased radiation can lead to a drop in the thermal pressure on the axis, and a radial collapse of laboratory jets increasing the collimation. Although the initial data are consitent with the the idea that more radiative jets are better collimated, firm conclusions cannot yet be reached. The reason for the large radius of the aluminium flows may be because these cones were too thin, and were exploded by the laser. In additions, the velocity of the aluminium jets is much higher than for copper or gold. This means that at the same experimental delay the aluminium jets have reached a later stage in their evolution, and that simple comparison of images taken at the same delay is not suitable for meaningful conclusions to be drawn. Future experiments will adjust the target design to conserve a constant areal mass density for all materials - by increasing the thickness of the lighter materials - in an attempt to create jets with similar expansion speeds. In addition, the time at which images are recorded will be normalised to the jet velocity for each material, effectively comparing jets of the same length, in order to record a snapshot of the jets at the same point in their evolution.

The time-resolved self-emission diagnostics shows that the aluminium jets move around 3.5 times faster than the copper and gold jets. It is therefore expected that the aluminium jets are $~3.5$ longer for a given time delay. This is not seen to be the case in the self-emission images taken at 50 ns and 70 ns. The reason for this is that as the jet expands the density and temperature drop, and the levels of emission are too low to be detected with the current setup. This effect is less significant for the copper and gold jets - they are denser than aluminium jets, and the increased atomic number of these jets results in stronger emission than for aluminium. The intereferometry displays a fringe shift for the aluminium jets upto (and likely beyond) the field of view of the diagnostic, indicating that the aluminium jets are indeed substantially longer than copper and gold jets. The view window of the interferometry diagnostic would have to be increased in order to see the leading edge of the flow, and provide a more accurate measure of the jet length than that extracted from the self-emission images. 

\section{Conclusions}

Thin conical targets of aluminium, copper and gold have been irradiated with an intense laser. The resulting flows have been studied with a variety of optical diagnostics and x-ray radiography, and jets are seen to form. The preliminary findings from these experiments support the conclusion that jets that radiate more efficiently are better collimated, due to the subsequent decrease in the internal thermal pressure and a collapse on axis. Future experiments will investigate a reasonable way to compare jets of different materials, since given the dissparate propagation velocities a simple comparison at some fixed time delay is not appropriate. In addition an ambient medium will be introduced into the jet propagation region to simulate the interstellar medium, and to increase the astrophysical relevence of this work.

\acknowledgments
The authors wish to thank the target preparation group at the Rutherford-Appleton Laboratory and Gabriel Schaumann of the Technical University of Darmstadt for fabricating the cone targes. CDG acknowledges financial support from R\'{e}gion Ile-de-France and RTRA.


\begin{thebibliography}{}
\ifx \bisbn   \undefined \def \bisbn  #1{ISBN #1}   \fi
\ifx \binits  \undefined \def \binits#1{#1} \fi
\ifx \bauthor  \undefined \def \bauthor#1{#1} \fi
\ifx \batitle  \undefined \def \batitle#1{#1} \fi
\ifx \bjtitle  \undefined \def \bjtitle#1{#1} \fi
\ifx \bvolume  \undefined \def \bvolume#1{\textbf {#1}} \fi
\ifx \byear  \undefined \def \byear#1{#1} \fi
\ifx \bissue  \undefined \def \bissue#1{#1} \fi
\ifx \bfpage  \undefined \def \bfpage#1{#1} \fi
\ifx \blpage  \undefined \def \blpage #1{#1} \fi
\ifx \burl  \undefined \def \burl#1{#1} \fi
\ifx \doiurl  \undefined \def \doiurl#1{#1} \fi
\ifx \betal  \undefined \def \betal#1{#1} \fi
\ifx \binstitutionaled  \undefined \def \binstitutionaled#1{#1} \fi
\ifx \binstitute  \undefined \def \binstitute#1{#1} \fi
\ifx \bctitle  \undefined \def \bctitle#1{#1} \fi
\ifx \beditor  \undefined \def \beditor#1{#1} \fi
\ifx \bpublisher  \undefined \def \bpublisher#1{#1} \fi
\ifx \bbtitle  \undefined \def \bbtitle#1{#1} \fi
\ifx \bedition  \undefined \def \bedition#1{#1} \fi
\ifx \bseriesno  \undefined \def \bseriesno#1{#1} \fi
\ifx \blocation  \undefined \def \blocation#1{#1} \fi
\ifx \bsertitle  \undefined \def \bsertitle#1{#1} \fi
\ifx \bsnm \undefined \def \bsnm#1{#1} \fi
\ifx \bsuffix \undefined \def \bsuffix#1{#1} \fi
\ifx \bparticle \undefined \def \bparticle#1{#1} \fi
\ifx \barticle \undefined \def \barticle#1{#1} \fi
\ifx \botherref \undefined \def \botherref #1{#1} \fi
\ifx \url \undefined \def \url#1{#1} \fi
\ifx \bchapter \undefined \def \bchapter#1{#1} \fi
\ifx \bbook \undefined \def \bbook#1{#1} \fi
\ifx \bcomment \undefined \def \bcomment#1{#1} \fi
\ifx \oauthor \undefined \def \oauthor#1{#1} \fi
\def \endbibitem {}

\bibitem[\protect\citeauthoryear{Ampleford et~al.}{2005}]{amp:for05}
\begin{barticle}
\bauthor{\bsnm{Ampleford},~\binits{D.J.}},
  \bauthor{\bsnm{Lebedev},~\binits{S.V.}},
  \bauthor{\bsnm{Ciardi},~\binits{A.}}, \oauthor{\bsnm{Bland},~\binits{S.N.}},
  \bauthor{\bsnm{Bott},~\binits{S.C.}},
  \bauthor{\bsnm{Chittenden},~\binits{J.P.}},
  \bauthor{\bsnm{Hall},~\binits{G.}}, \oauthor{\bsnm{Jennings},~\binits{C.A.}},
  \bauthor{\bsnm{Armitage},~\binits{J.}}, \oauthor{\bsnm{Blyth},~\binits{G.}},
  \bauthor{\bsnm{Christie},~\binits{S.}},
  \bauthor{\bsnm{Rutland},~\binits{L.}}:
\bjtitle{Astrophys. Space Sci.}  \bvolume{298}, \bfpage{241}(\byear{2005})
\end{barticle}
\endbibitem

\bibitem[\protect\citeauthoryear{Foster et~al.}{2005}]{fos:hig05}
\begin{barticle}
\bauthor{\bsnm{Foster},~\binits{J.M.}}, \bauthor{\bsnm{Wilde},~\binits{B.H.}},
  \bauthor{\bsnm{Rosen},~\binits{P.A.}},
  \bauthor{\bsnm{Williams},~\binits{R.J.R.}},
  \bauthor{\bsnm{Blue},~\binits{B.E.}}, \bauthor{\bsnm{Coker},~\binits{R.F.}},
  \bauthor{\bsnm{Drake},~\binits{R.P.}}, \bauthor{\bsnm{Frank},~\binits{A.}},
  \bauthor{\bsnm{Keiter},~\binits{P.A.}},
  \bauthor{\bsnm{Khokhlov},~\binits{A.M.}},
  \bauthor{\bsnm{Knauer},~\binits{J.P.}},
  \bauthor{\bsnm{Perry},~\binits{T.S.}}:
\bjtitle{Astrophys. J. Lett.} \bvolume{634}, \bfpage{L77}
  (\byear{2005})
\end{barticle}
\endbibitem

\bibitem[\protect\citeauthoryear{Gregory et~al.}{2008}]{gre:ast08}
\begin{barticle}
\bauthor{\bsnm{Gregory},~\binits{C.D.}}, \bauthor{\bsnm{Howe},~\binits{J.}},
  \bauthor{\bsnm{Loupias},~\binits{B.}}, \bauthor{\bsnm{Myers},~\binits{S.}},
  \bauthor{\bsnm{Notley},~\binits{M.M.}}, \bauthor{\bsnm{Sakawa},~\binits{Y.}},
  \bauthor{\bsnm{Oya},~\binits{A.}}, \bauthor{\bsnm{Kodama},~\binits{R.}},
  \bauthor{\bsnm{Koenig},~\binits{M.}},
  \bauthor{\bsnm{Woolsey},~\binits{N.C.}}:
\bjtitle{Astrophys. J.} \bvolume{676}, \bfpage{420} (\byear{2008})
\end{barticle}
\endbibitem

\bibitem[\protect\citeauthoryear{Gregory et~al.}{2008b}]{gre:ast08b}
\begin{barticle}
\bauthor{\bsnm{Gregory},~\binits{C.D.}}, \bauthor{\bsnm{Loupias},~\binits{B.}},
  \bauthor{\bsnm{Waugh},~\binits{J.}}, \bauthor{\bsnm{Barroso},~\binits{P.}},
  \bauthor{\bsnm{Bouquet},~\binits{S.}},
  \bauthor{\bsnm{Brambrink},~\binits{E.}}, \bauthor{\bsnm{Dono},~\binits{S.}},
  \bauthor{\bsnm{Falize},~\binits{E.}}, \bauthor{\bsnm{Howe},~\binits{J.}},
  \bauthor{\bsnm{Kuramitsu},~\binits{Y.}},
  \bauthor{\bsnm{Kodama},~\binits{R.}}, \bauthor{\bsnm{Koenig},~\binits{M.}},
  \bauthor{\bsnm{Michaut},~\binits{C.}}, \bauthor{\bsnm{Myers},~\binits{S.}},
  \bauthor{\bsnm{Nazarov},~\binits{W.}},
  \bauthor{\bsnm{Notley},~\binits{M.M.}}, \bauthor{\bsnm{Oya},~\binits{A.}},
  \bauthor{\bsnm{Pikuz},~\binits{S.}},
  \bauthor{\bparticle{le~}\bsnm{Gloahec},~\binits{M.R.}},
  \bauthor{\bsnm{Sakawa},~\binits{Y.}}, \bauthor{\bsnm{Spindloe},~\binits{C.}},
  \bauthor{\bsnm{Streeter},~\binits{M.}},
  \bauthor{\bsnm{Wilson},~\binits{L.A.}},
  \bauthor{\bsnm{Woolsey},~\binits{N.C.}}:
\bjtitle{Plasma Phys. and Cont. Fusion} \bvolume{50}, \bfpage{124039}
  (\byear{2008b})
\end{barticle}
\endbibitem

\bibitem[\protect\citeauthoryear{Lebedev et~al.}{2004}]{leb:jet04}
\begin{barticle}
\bauthor{\bsnm{Lebedev},~\binits{S.V.}},
  \bauthor{\bsnm{Ampleford},~\binits{D.}},
  \bauthor{\bsnm{Ciardi},~\binits{A.}}, \bauthor{\bsnm{Bland},~\binits{S.N.}},
  \bauthor{\bsnm{Chittenden},~\binits{J.P.}},
  \bauthor{\bsnm{Haines},~\binits{M.G.}}, \bauthor{\bsnm{Frank},~\binits{A.}},
  \bauthor{\bsnm{Blackman},~\binits{E.G.}},
  \bauthor{\bsnm{Cunningham},~\binits{A.}}:
\bjtitle{Astrophys. J.} \bvolume{616}, \bfpage{988} (\byear{2004})
\end{barticle}
\endbibitem

\bibitem[\protect\citeauthoryear{Lebedev et~al.}{2002}]{leb:lab02}
\begin{barticle}
\bauthor{\bsnm{Lebedev},~\binits{S.V.}},
  \bauthor{\bsnm{Chittenden},~\binits{J.P.}},
  \bauthor{\bsnm{Beg},~\binits{F.N.}}, \bauthor{\bsnm{Bland},~\binits{S.N.}},
  \bauthor{\bsnm{Ciardi},~\binits{A.}},
  \bauthor{\bsnm{Ampleford},~\binits{D.}},
  \bauthor{\bsnm{Hughes},~\binits{S.}}, \bauthor{\bsnm{Haines},~\binits{M.G.}},
  \bauthor{\bsnm{Frank},~\binits{A.}},
  \bauthor{\bsnm{Blackman},~\binits{E.G.}},
  \bauthor{\bsnm{Gardiner},~\binits{T.}}:
\bjtitle{Astrophys. J.} \bvolume{564}, \bfpage{113} (\byear{2002})
\end{barticle}
\endbibitem

\bibitem[\protect\citeauthoryear{Loupias et~al.}{2007}]{lou:sup07}
\begin{barticle}
\bauthor{\bsnm{Loupias},~\binits{B.}}, \bauthor{\bsnm{Koenig},~\binits{M.}},
  \bauthor{\bsnm{Falize},~\binits{E.}}, \bauthor{\bsnm{Bouquet},~\binits{S.}},
  \bauthor{\bsnm{Ozaki},~\binits{N.}},
  \bauthor{\bsnm{Benuzzi-Mounaix},~\binits{A.}},
  \bauthor{\bsnm{Vinci},~\binits{T.}}, \bauthor{\bsnm{Michaut},~\binits{C.}},
  \bauthor{\bparticle{le~}\bsnm{Gloahec},~\binits{M.R.}},
  \bauthor{\bsnm{Nazarov},~\binits{W.}},
  \bauthor{\bsnm{Courtois},~\binits{C.}},
  \bauthor{\bsnm{Aglitsky},~\binits{Y.}},
  \bauthor{\bsnm{Faenov},~\binits{A.Y.}}, \bauthor{\bsnm{Pikuz},~\binits{T.}}:
\bjtitle{Phys. Rev. Lett.} \bvolume{99}, \bfpage{265001} (\byear{2007})
\end{barticle}
\endbibitem

\bibitem[\protect\citeauthoryear{Mizuta, Yamada \& Takabe}{2002}]{miz:num02}
\begin{barticle}
\bauthor{\bsnm{Mizuta},~\binits{A.}}, \bauthor{\bsnm{Yamada},~\binits{S.}},
  \bauthor{\bsnm{Takabe},~\binits{H.}}:
\bjtitle{Astrophys. J.} \bvolume{57}, \bfpage{635} (\byear{2002})
\end{barticle}
\endbibitem

\bibitem[\protect\citeauthoryear{Nicola\"i et~al.}{2008}]{nic:stu08}
\begin{barticle}
\bauthor{\bsnm{Nicola\"i},~\binits{P.}}, \bauthor{\bsnm{Stenz},~\binits{C.}},
  \bauthor{\bsnm{Kasperczuk},~\binits{A.}},
  \bauthor{\bsnm{Pisarczyk},~\binits{T.}}, \bauthor{\bsnm{Klir},~\binits{D.}},
  \bauthor{\bsnm{Juha},~\binits{L.}}, \bauthor{\bsnm{Krousky},~\binits{E.}},
  \bauthor{\bsnm{Masek},~\binits{K.}}, \bauthor{\bsnm{Pfeifer},~\binits{M.}},
  \bauthor{\bsnm{Rohlena},~\binits{K.}}, \bauthor{\bsnm{Skala},~\binits{J.}},
  \bauthor{\bsnm{Tikhonchuk},~\binits{V.}},
  \bauthor{\bsnm{Ribeyre},~\binits{X.}}, \bauthor{\bsnm{Galera},~\binits{S.}},
  \bauthor{\bsnm{Schurtz},~\binits{G.}},
  \bauthor{\bsnm{Ullschmied},~\binits{J.}},
  \bauthor{\bsnm{Kalal},~\binits{M.}}, \bauthor{\bsnm{Kravarik},~\binits{J.}},
  \bauthor{\bsnm{Kubes},~\binits{P.}}, \bauthor{\bsnm{Pisarczyk},~\binits{P.}},
  \bauthor{\bsnm{Schlegel},~\binits{T.}}:
\bjtitle{Phys. Plasmas} \bvolume{15}, \bfpage{082701} (\byear{2008})
\end{barticle}
\endbibitem

\bibitem[\protect\citeauthoryear{Purvis et~al.}{2010}]{pur:col10}
\begin{barticle}
\bauthor{\bsnm{Purvis},~\binits{M.A.}}, \bauthor{\bsnm{Grava},~\binits{J.}},
  \bauthor{\bsnm{Filevich},~\binits{J.}}, \bauthor{\bsnm{Ryan},~\binits{D.P.}},
  \bauthor{\bsnm{Moon},~\binits{S.J.}}, \bauthor{\bsnm{Dunn},~\binits{J.}},
  \bauthor{\bsnm{Shlyaptsev},~\binits{V.N.}},
  \bauthor{\bsnm{Rocca},~\binits{J.J.}}:
\bjtitle{Phys. Rev. E} \bvolume{81}, \bfpage{036408} (\byear{2010})
\end{barticle}
\endbibitem

\bibitem[\protect\citeauthoryear{Reipurth and Bally}{2001}]{rei:her01}
\begin{barticle}
\bauthor{\bsnm{Reipurth},~\binits{B.}}, \bauthor{\bsnm{Bally},~\binits{J.}}:
\bjtitle{Annu. Review Astron. Astrophys.} \bvolume{39},
  \bfpage{403} (\byear{2001})
\end{barticle}
\endbibitem

\bibitem[\protect\citeauthoryear{Shigemori et~al.}{2000}]{shi:exp00}
\begin{barticle}
\bauthor{\bsnm{Shigemori},~\binits{K.}}, \bauthor{\bsnm{Kodama},~\binits{R.}},
  \bauthor{\bsnm{Farley},~\binits{D.R.}}, \bauthor{\bsnm{Koase},~\binits{T.}},
  \bauthor{\bsnm{Estbrook},~\binits{K.G.}},
  \bauthor{\bsnm{Remington},~\binits{B.A.}},
  \bauthor{\bsnm{Ryutov},~\binits{D.D.}}, \bauthor{\bsnm{Ochi},~\binits{Y.}},
  \bauthor{\bsnm{Azechi},~\binits{H.}}, \bauthor{\bsnm{Stone},~\binits{J.}},
  \bauthor{\bsnm{Turner},~\binits{N.}}:
\bjtitle{Phys. Rev. E} \bvolume{62}, \bfpage{8838}
  (\byear{2000})
\end{barticle}
\endbibitem

\bibitem[\protect\citeauthoryear{Waugh et~al.}{2009}]{wau:aje09}
\begin{barticle}
\bauthor{\bsnm{Waugh},~\binits{J.N.}}, \oauthor{\bsnm{Gregory},~\binits{C.D.}},
  \bauthor{\bsnm{Wilson},~\binits{L.A.}},
  \bauthor{\bsnm{Loupias},~\binits{B.}},
  \bauthor{\bsnm{Brambrink},~\binits{E.}},
  \bauthor{\bsnm{Koenig},~\binits{M.}}, \oauthor{\bsnm{Sakawa},~\binits{Y.}},
  \bauthor{\bsnm{Kuramitsu},~\binits{Y.}},
  \bauthor{\bsnm{Takabe},~\binits{H.}}, \oauthor{\bsnm{Kodama},~\binits{R.}},
  \bauthor{\bsnm{Woolsey},~\binits{N.C.}}:
\bjtitle{Astrophys. Space Sci.}  \bvolume{322}, \bfpage{31}(\byear{2009})
\end{barticle}
\endbibitem

\end{thebibliography}
\end{document}